\documentclass{PoS}

\usepackage{psfig}

\title{Magnetic Moment of Vector Mesons in the Background Field Method}

\ShortTitle{Magnetic Moment of Vector Mesons}

\author{\speaker{Frank X. Lee} and Scott Moerschbacher \\
Physics Department,
The George Washington University, Washington, DC 20052, USA \\
E-Mail: \email{fxlee@gwu.edu}}
\author{Walter Wilcox \\
Department of Physics, Baylor University, Waco, TX 76798, USA}

\abstract{We report some results for the magnetic moments of vector mesons extracted
from mass shifts in the presence of static external magnetic fields.
The calculations are done on $24^4$ quenched lattices using standard Wilson actions,
with $\beta$=6.0 and pion mass down to 500 MeV.
The results are compared to those from the form factor method.}

\FullConference{The XXV International Symposium on Lattice Field Theory\\
                 July 30 - August 4 2007\\
                 Regensburg, Germany}

\begin{document}

 \section{Introduction}
Magnetic moment is a fundamental property of particles. It
determines the dynamical response of a bound system to a soft external stimulus, 
and provides valuable insight into internal strong interaction structure. 
Efforts to compute the magnetic moment on the lattice come in two categories. 
One is the form factor method which involves 
three-point functions~\cite{mart89,leinweber3,wilcox92,ji02,horse03,zanotti04,cloet04}. 
The other is the background field method using only two-point 
functions (mass shifts)~\cite{mart82,bernard82,smit87,rubin95}.
The form factor method requires an extrapolation to zero momentum transfer $G_M(Q^2=0)$
due to the non-vanishing minimum discrete momentum on the lattice~\cite{wilcox02}. 
The background field method, on the other hand, accesses the magnetic moment directly 
and cleanly but is limited to static properties due to the use of a static field.
Here we report a calculation of the vector meson magnetic moments in this method, 
in parallel to a recent calculation in the form factor method~\cite{Hedditch07}.
It is an extension of our earlier work on baryon magnetic moments~\cite{Lee05} and 
electric~\cite{Joe05} and magnetic polarizabilities~\cite{Lee06} in the same method.

 \section{Method}
For a Dirac particle of spin $s$ in uniform fields, 
\begin{equation}
E_\pm=m\pm\mu B
\end{equation}
where the upper sign means spin up and the lower sign means spin-down relative 
to the magnetic field, and 
$\mu=g {e\over 2m}s$. 
We use the following method to extract the g factors,
\begin{equation}
g=\left( {2m_+ m_- \over m_+ + m_-} \right){(E_+ - m_+)-(E_- - m_-)\over eBs}.
\label{g1}
\end{equation}

In order to place a magnetic field on the lattice, we construct an 
analogy to the continuum case. The fermion action is modified 
by the minimal coupling prescription 
\begin{equation}
D_\mu = \partial_\mu+gG_\mu + q A_\mu
\end{equation}
where $q$ is the charge of the fermion field and $A_\mu$ is the vector 
potential describing the background field. On the lattice, the prescription
amounts to multiplying a U(1) phase factor to the gauge links.
Choosing $A_y = B x $, a constant magnetic field B can be introduced 
in the $z$-direction. Then the phase factor is in the y-links
\begin{equation}
U_y \rightarrow \exp{(iqa^2Bx)} U_y.
\end{equation}
The computational demand can be divided into three categories.
The first is a {\em fully-dynamical} calculation. For each value of external field, a new dynamical ensemble is needed that couples to u-quark (q=1/3), d-and s-quark (q=-2/3). This requires a Monte Carlo algorithm that can treat the three flavors separately. Quark propagators are then computed on the ensembles with matching values. This has not been attempted.
The second can be termed {\em re-weighting} in which a perturbative expansion of action in terms of external field is performed (see Ref.~\cite{Engel07} for a calculation of the neutron electric polarizability in this method).
The third is {\em U(1) quenched}. No field is applied in the Monte-Carlo generation of the gauge fields, only in the valence quark propagation in the given gauge background.
In this case, any gauge ensemble can be used to compute valence quark propagators.

We use standard Wilson actions on $24^4$ lattice at $\beta=6.0$, both SU(3) and U(1) quenched, and six kappa values $\kappa$=0.1515, 0.1525, 0.1535, 
0.1540, 0.1545, 0.1555, corresponding to pion mass of 1015, 908, 794, 732, 667, 522 MeV.
The critical value of kappa is $\kappa_c$=0.1571.
The strange quark mass is set at $\kappa$=0.1535. The source location for the quark propagators 
is (x,y,z,t)=(12,1,1,2).
We analyzed 87 configurations.
The following five dimensionless numbers 
$\eta=qBa^2$=+0.00036, -0.00072, +0.00144, -0.00288, +0.00576 give four small B fields 
(two positive, two negative) at 
$eBa^2$=-0.00108, +0.00216,  -0.00432, +0.00864 for both u and d (or s) quarks. 
These field values do not obey the quantization condition for periodicity since they cause too large a distortion to the system. 
To minimize the boundary effects, we work with fixed (or Dirichlet) b.c. in the x-direction and 
large $N_x$, so that quarks originating in the middle of the lattice have little chance of 
propagating to the edge. 
To eliminate the contamination from the even-power terms, we calculate mass shifts 
both in the field $B$ and its reverse $-B$ for each value of $B$, 
then take the difference and divide by 2.
Another benefit of repeating the calculation with the field reversed is that 
by taking the average of $\delta m (B)$ and $\delta m (-B)$ in the same dataset, 
one can eliminate the odd-powered terms in the mass shift. 
The coefficient of the leading quadratic term 
is directly related to the magnetic polarizability~\cite{Lee06}.

\section{Interpolating field}
For the $\rho^+$ meson, we use the polarized forms
\begin{equation}
\eta_\pm={1\over \sqrt{2}}\bar{d}\left(\mp\gamma_x - i\gamma_y \right) u
= {1\over \sqrt{2}}\left(\eta_x \pm i\eta_y \right)
\end{equation}
The interaction energies $E_\pm$ are extracted from the correlation functions
\begin{equation}
\langle \bar{\eta_\pm} \eta_\pm \rangle = 
{1\over 2}\left[ \langle \bar{\eta_x} \eta_x \rangle
    \pm i \left( \langle \bar{\eta_x} \eta_y \rangle 
                -\langle \bar{\eta_y} \eta_x \rangle \right)
                +\langle \bar{\eta_y} \eta_y \rangle \right].
\label{polar2}
\end{equation}
Eq.~(\ref{polar2}) implies that the polarization comes from the imaginary parts of 
the off-diagonal correlation between x and y components. 
These imaginary parts are zero in the absence of the external field.
Other vector mesons have similar forms with different quark content $\rho^-=\bar{u}d$, 
$\rho^0=\bar{u}u-\bar{d}d$, $\phi=\bar{s}s$, $K^{*+}=\bar{s}u$, $K^{*-}=\bar{u}s$, 
$K^{*0}=\bar{d}s-\bar{s}d$. By symmetry, the magnetic moments are expected to be related 
by 
\begin{equation}
\mu_{\rho^-}=-\mu_{\rho^+},\hspace{2mm} \mu_{\rho^0}=0; 
\end{equation}
and
\begin{equation}
\mu_{K^{*-}}=-\mu_{K^{*+}},\hspace{2mm} \mu_{K^{*0}}\;\; \mbox{small}. 
\end{equation}
These relations are borne out numerically in our calculations.

%
\begin{figure}
\parbox{.5\textwidth}{%
\psfig{file=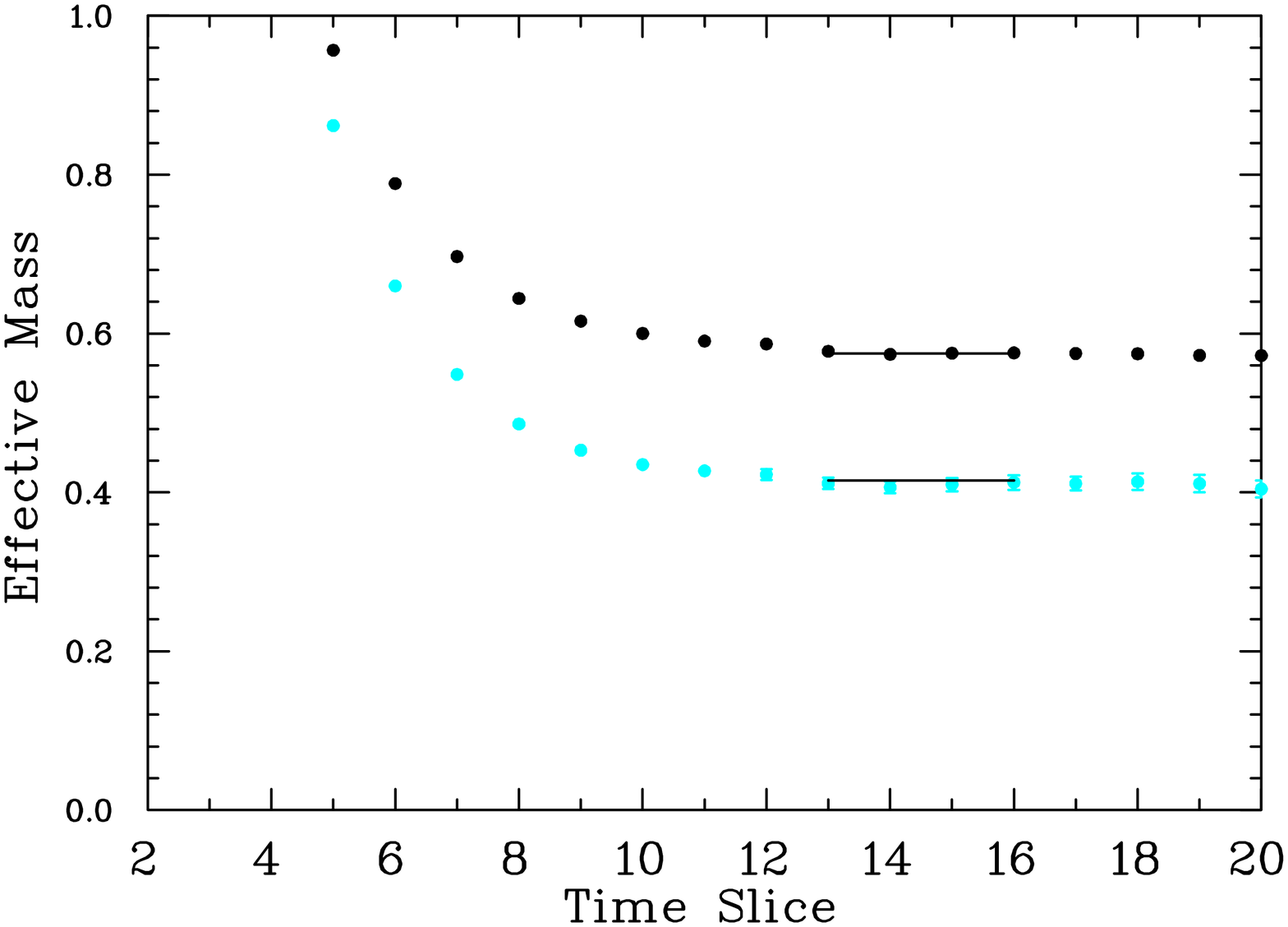,width=3.0in,angle=90}}
\parbox{.5\textwidth}{%
\psfig{file=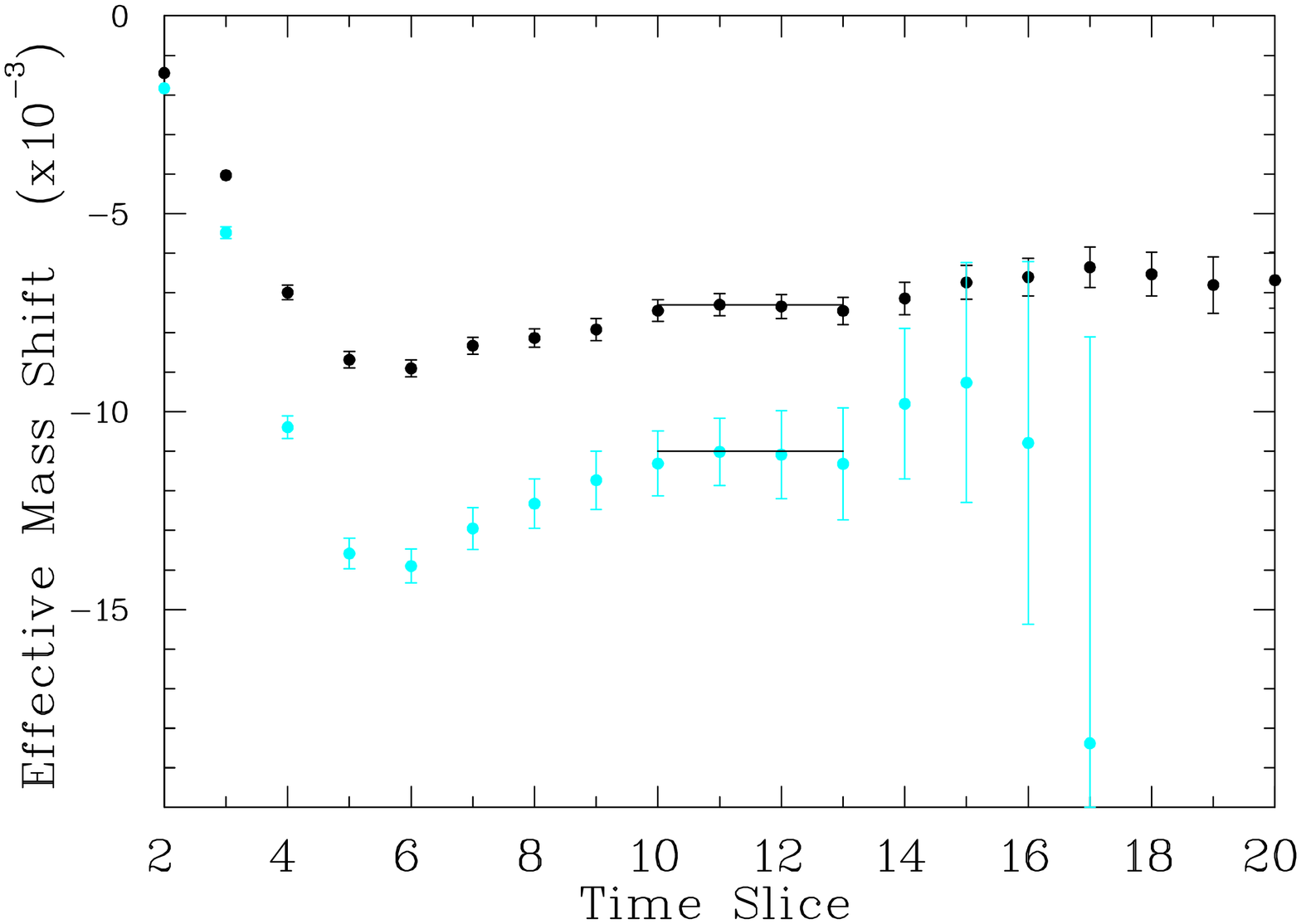,width=3.0in,angle=90}}
\caption{Effective mass plot for the $\rho^+$ mass (left) 
at zero field and mass shifts (right) at the weakest magnetic field
in lattice units, corresponding to the heaviest and lightest quark masses.}
\label{emass2}
\end{figure}
%
%
\begin{figure}
\centerline{\psfig{file=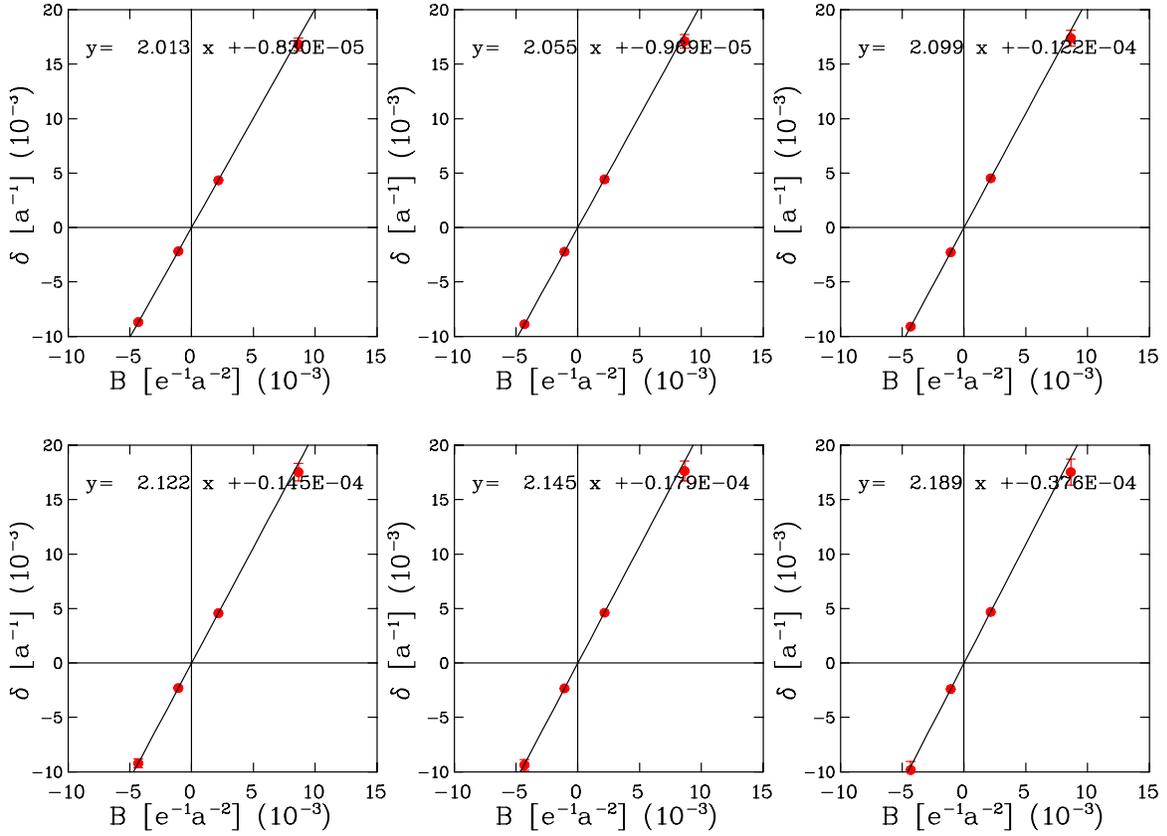,width=6.0in,angle=90}}
\caption{Mass shifts for the $\rho^+$ meson as a function of the magnetic field 
in lattice units at the six quark masses (lightest in the lower right corner). 
The slope of the mass shift at each quark mass gives the g factor corresponding to 
that quark mass.
The line is a fit using only the two smallest B values.}
\label{rhop_shift_linear}
\end{figure}
%

 \section{Results and discussion}
Fig.~\ref{emass2} displays a typical effective mass plot for both the mass and the mass shifts.
Good plateaus exist for all six quark masses.
Our results are extracted from the time window 10 to 13.
Fig.~\ref{rhop_shift_linear} shows the mass shifts, 
defined as $\delta=g (eBs)$ from Eq.~(\ref{g1}), as a function of the field
for the $\rho^+$ meson. The slope gives the g-factor.
There is good linear behavior going through the origin for all the fields 
when the quark mass is heavy, 
an indication that contamination from the higher-power terms has been 
effectively eliminated by the $(\delta(B)-\delta(-B))/2$ procedure. 
This is also confirmed numerically by the smallness of 
intercept as shown in the same figure. 
At the lightest quark mass (lower right corner), there is a slight deviation from linear behavior at the stronger fields.
For this reason, we only use the two smallest field values 
to do the linear fit at all the quark masses.

\begin{figure}
\parbox{.5\textwidth}{%
\psfig{file=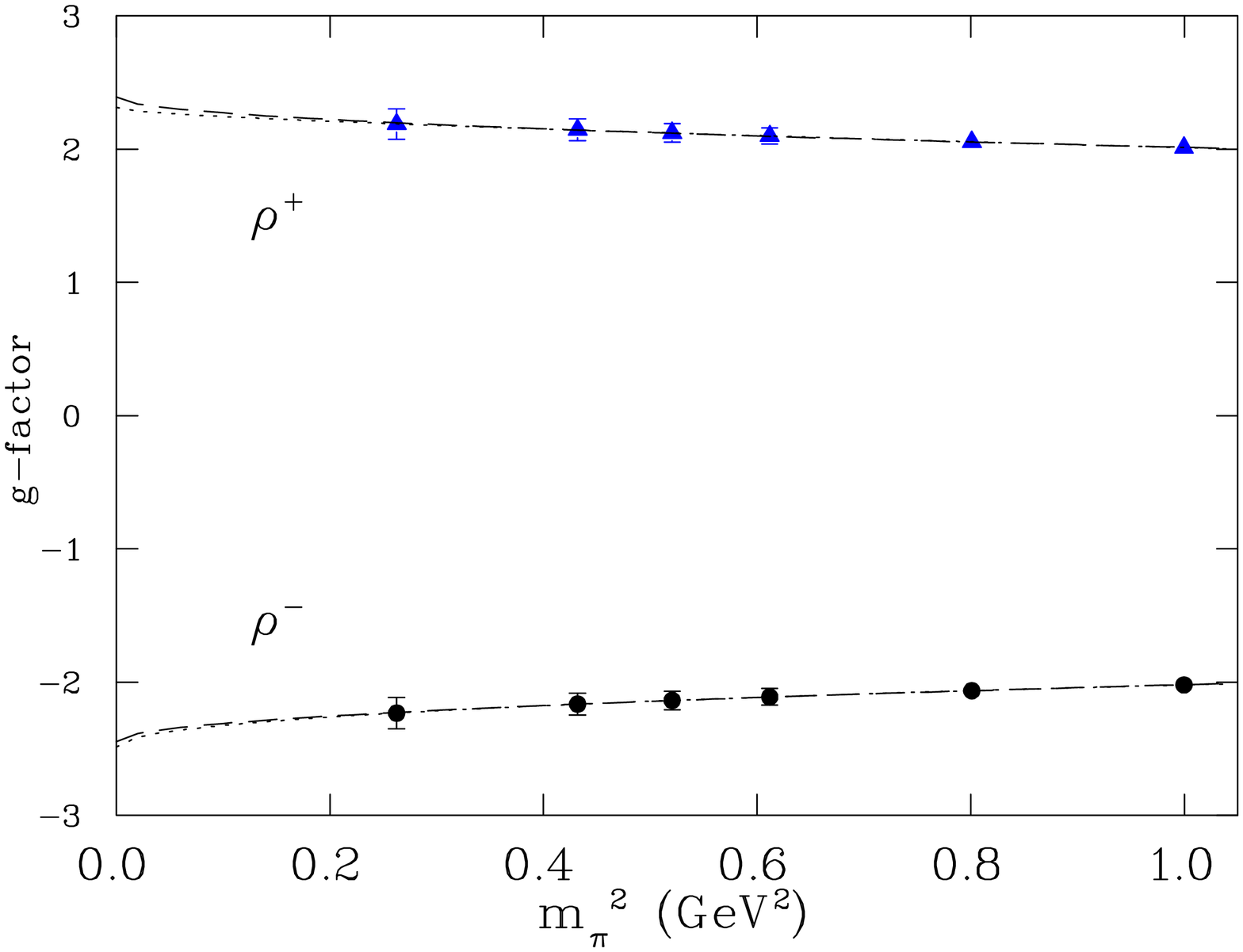,width=3.0in,angle=90}}
\parbox{.5\textwidth}{%
\psfig{file=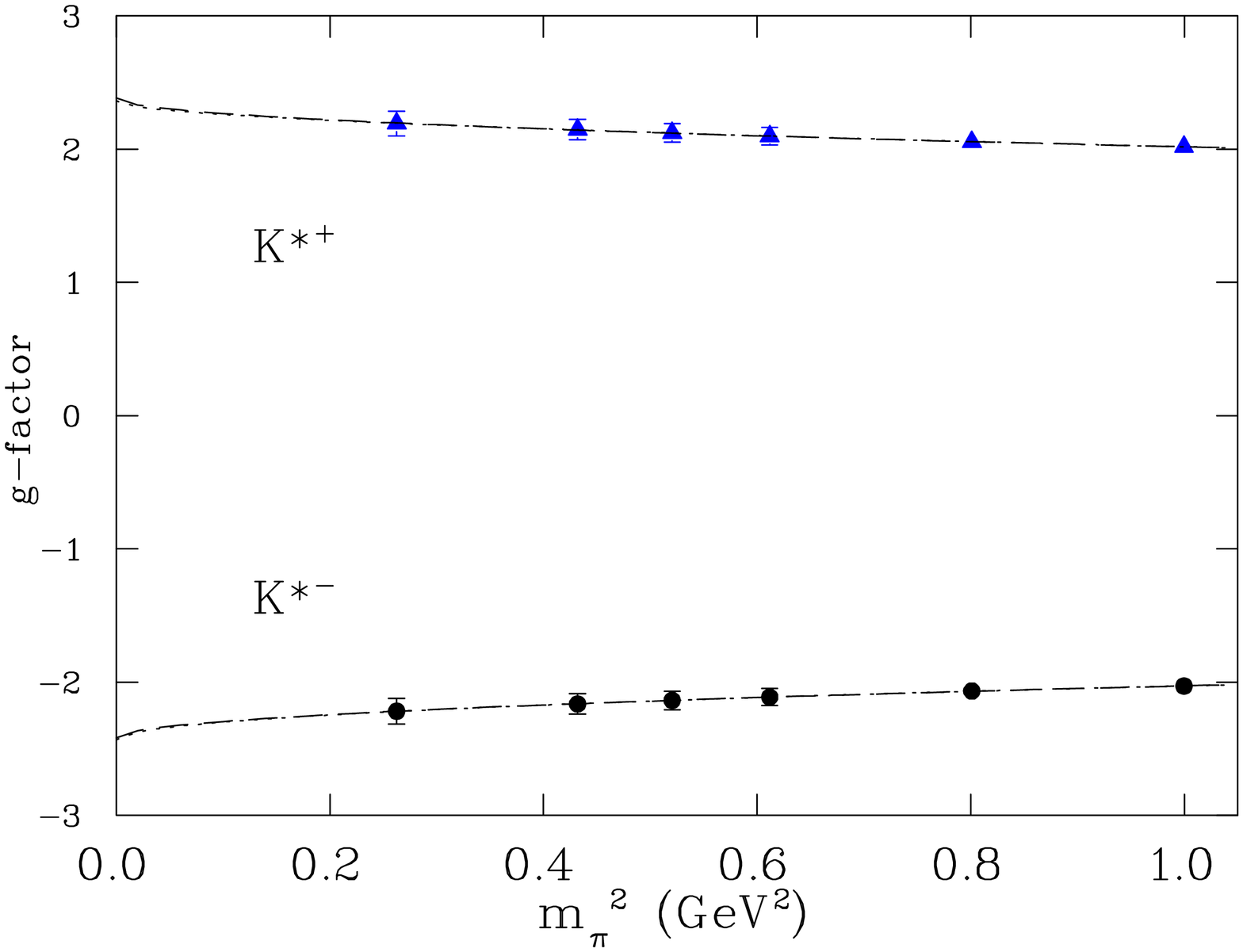,width=3.0in,angle=90}}
\caption{G-factors for the $\rho^\pm$ (left) and $K^*$ (right) mesons 
as a function of pion mass squared. 
The 2 lines are chiral fits according to
Eq.~(\protect\ref{chiral1}) (dashed), Eq.~(\protect\ref{chiral2}) (dotted).}
\label{gfac_vector}
\end{figure}

Fig.~\ref{gfac_vector} shows the g-factors for the vector mesons as a function
of pion mass squared.
The lines are simple chiral fits using the ansatzs
\begin{equation}
g=a_0 + a_1 m_\pi,
\label{chiral1}
\end{equation}
and
\begin{equation}
g=a_0 + a_1 m_\pi+a_2 m_\pi^2.
\label{chiral2}
\end{equation}
They serve to show that there is onset of non-analytic behavior as pion mass is lowered, 
so a linear extrapolation is probably not a good idea. 
But overall the g-factors have a fairly weak quark mass dependence. At large quark masses, 
the g-factor of $\rho^+$ approaches 2, consistent with a previous lattice calculation 
using charge-overlap method~\cite{Wilcox97}.
Our results for $\rho^+$ are slightly higher than those from the form factor 
method (see Fig.8 in~\cite{Hedditch07}). 
The results confirmed that $g_{\rho^-}=-g_{\rho^+}$ and $g_{K^{*-}}=-g_{K^{*+}}$.
We confirmed $g_{\rho^0}=0$ numerically (not shown). The results also show that as far as g-factors are concerned 
the $\rho$ mesons are quite similar to their strange counterparts ${K^{*}}$ mesons.

\begin{figure}
\centerline{\psfig{file=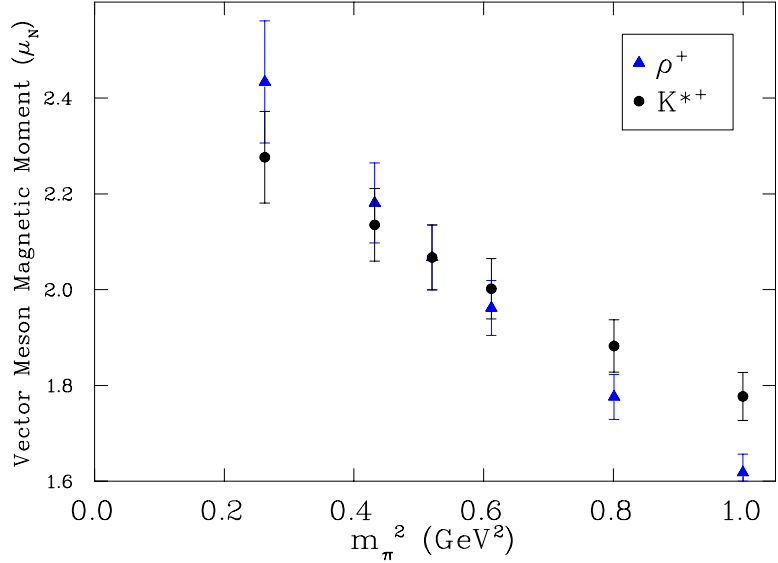,width=4.0in,angle=90}}
\caption{Magnetic moments (in nuclear magnetons) for $\rho^+$ and $K^{*+}$.}
\label{mag_vector}
\end{figure}

Note that the extracted g-factors are in the particle's natural 
magnetons.  To convert them into magnetic moments in terms of the commonly-used nuclear 
magnetons ($\mu_N$), we need to scale
the results by the factor $938/M$ where $M$ is the mass of the particle
measured in the same calculation at each quark mass.
Fig.~\ref{mag_vector} shows the results for ${\rho^+}$ and ${K^{*+}}$.
The different quark-mass dependence between ${\rho^+}$ and ${K^{*+}}$ mostly 
comes from that in their masses that are used to convert the g-factors to magnetic moments.
The values at the chiral limit extrapolated from Eq.~(\ref{g1}) are 
$\mu_{\rho^+}=3.25(3) \mu_N$ and $\mu_{K^{*+}}=2.81(1) \mu_N$. 
There is no experimental information.
Compared to the form factor method (see Fig.7 in~\cite{Hedditch07}), our results are again 
a little higher.
At the strange quark mass point (the 3rd data point from the left), the two coincide to
give a prediction for the magnetic moment of the $\phi(1020)$ meson, $\mu_\phi=2.07(7) \mu_N$.

%
%
%
%

\begin{figure}
\centerline{\psfig{file=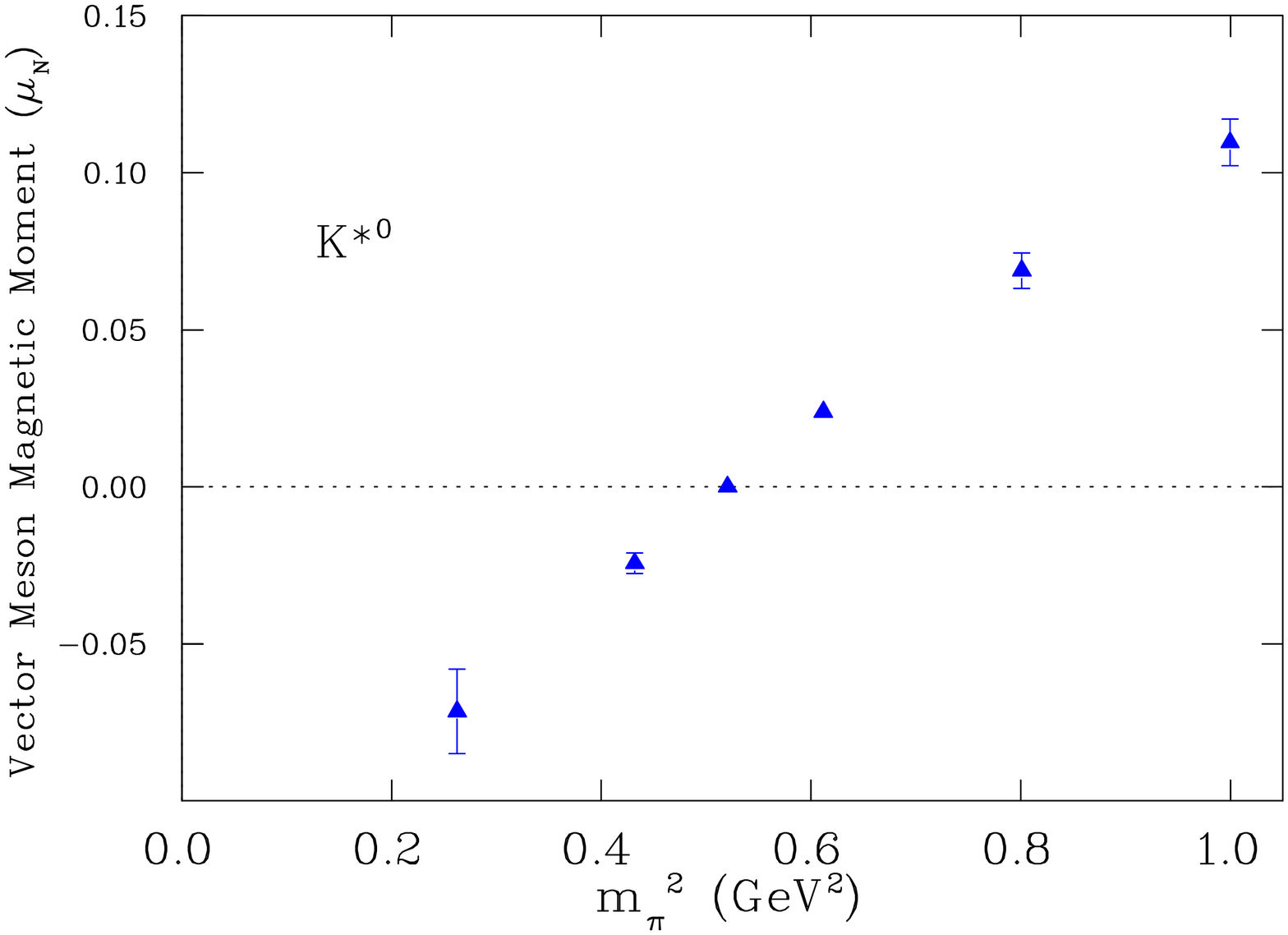,width=4.0in,angle=90}}
\caption{Magnetic moments (in nuclear magnetons) for $K^{*0}$.}
\label{mag_vector1}
\end{figure}

Fig.~\ref{mag_vector1} shows the results for ${K^{*0}}$.
Our results confirm the expectation that $\mu_{K^{*0}}$ is small.
It is positive when the d-quark is heavier than the s-quark, 
exactly zero when they are equal, and turns negative when the d-quark is lighter than the s-quark.
The same behavior has been observed in the form factor method (see Fig.11 in~\cite{Hedditch07}).

%
%

 \section{Conclusion}
In conclusion, we have computed the magnetic moment of vector mesons
on the lattice using the background field method and standard lattice 
technology. Our results are consistent with those from the form factor method.
There is no experimental information so the lattice results can serve as 
first-principles predictions.
The calculation can be improved by providing a full account of systematic errors 
present in the results, such as finite-volume effects. 
In addition,  there is a need to push the calculations to smaller pion masses 
so that reliable chiral extrapolations can be applied.
Nonetheless, our results demonstrate that the method is robust and relatively cheap. 
Only mass shifts are required. 
This may facilitate the push to smaller pion masses, perhaps with the help of chiral fermions 
(overlap, domain-wall, twisted mass, ...). 
Finally, we await fully-dynamical background-field calculations 
in order to see the effects of the quenched approximation.

\section*{Acknowledgment}
This work is supported in part by U.S. Department of Energy
under grant DE-FG02-95ER40907.  W.W. acknowledges a research leave from Baylor University.
The computing resources at NERSC and JLab have been used.


\begin{thebibliography}{99}

\bibitem{mart89} G. Martinelli and C.T. Sachrajda,
Nucl. Phys.  {\bf B316}, 355 (1989).

\bibitem{leinweber3} D.B. Leinweber, T. Draper, R.M. Woloshyn,
Phys. Rev. {\bf D43}, 1659 (1991);
Phys. Rev. {\bf D46}, 3067 (1992);
Phys. Rev. {\bf D48}, 2230 (1993).

\bibitem{wilcox92} W. Wilcox, T. Draper and K.F. Liu,
Phys. Rev. {\bf D46}, 1109 (1992).

\bibitem{ji02} V. Gadiyak, X. Ji, C. Jung,
Phys. Rev. {\bf D65}, 094510 (2002).

\bibitem{horse03} G\"{o}ckeler {\it et. al.}, 
Phys. Rev. {\bf D71}, 034508 (2005).

\bibitem{zanotti04} J. Zanotti, Boinepalli, D.B. Leinweber, A.W. Williams, J.B. Zhang,
hep-lat/0401029.

\bibitem{cloet04} I.C. Cloet, D.B. Leinweber, A.W. Thomas,
Phys. Lett. {\bf B563}, 157 (2003).



\bibitem{mart82} G. Martinelli {\it et. al.},
Phys. Lett. {\bf B116}, 434 (1982).

\bibitem{bernard82} C. Bernard, T. Draper, K. Olynyk,
Phys. Rev. Lett. {\bf 49}, 1076 (1982);
C. Bernard, T. Draper, K. Olynyk,  M. Rushton, Nucl. Phys.  {\bf B220}, 508 (1983).

\bibitem{smit87} J. Smit and J.C. Vink, 
Nucl. Phys.  {\bf B286}, 485 (1987).

\bibitem{rubin95} H.R. Rubinstein, S. Solomon and T. Wittlich,
Nucl. Phys.  {\bf B457}, 577 (1995).

\bibitem{wilcox02} W. Wilcox,
Phys. Rev. {\bf D66}, 017502 (2002).

\bibitem{Hedditch07} J.N. Hedditch, W. Kamleh, B.G. Lasscock, D.B. Leinweber, A.G. Williams, J.M. Zanotti,
Phys. Rev. {\bf D75}, 094504 (2007).

\bibitem{Lee05} F.X. Lee, R.Kelly, L. Zhou, W. Wilcox,
Phys. Lett. {\bf B627}, 71 (2005).

\bibitem{Joe05} J. Christensen, W. Wilcox, F.X. Lee, L. Zhou,
Phys. Rev. {\bf D72}, 034503 (2005).

\bibitem{Lee06} F.X. Lee, L. Zhou, W. Wilcox, J. Christensen, 
Phys. Rev. {\bf D73}, 034503 (2006).

\bibitem{Engel07} M. Engelhardt, these proceedings.

\bibitem{Wilcox97} W. Anderson and W. Wilcox,
Annals of Phys. {\bf 255}, 34 (1997).

\end{thebibliography}
\end{document}